# THE DAEMON KERNEL OF THE SUN

E.M.Drobyshevski

*A.F.Ioffe Physico-Technical Institute, Russian Academy of Sciences,
194021 St.Petersburg, Russia
(E-mail: emdrob@pop.ioffe.rssi.ru)*

*Abstract.* The daemon-stimulated proton decay is fully capable of providing an appreciable part of the Sun luminosity as well as nonelectron flavor component in the solar neutrino flux. This follows (1) from our experiments on detection of negative daemons in Earth-crossing orbits, which give $\Delta t_{ex} \sim 1$ µs for the decay time of a daemon-containing proton (Drobyshevski *et al*, 2001), and (2) from an estimate of the total number of daemons which could be captured by the Sun from the Galactic disk (up to $\sim 2.4 \cdot 10^{30}$) (Drobyshevski, 1996). Because of their huge mass ($\sim 3 \cdot 10^{-5}$ g), the captured daemons settle down to the Sun's center to form there a kernel a few cm in size. The Earth has also such a kernel (Drobyshevski, 2001, 2002). The outside protons diffuse gradually into the kernel to decay there with a release of energy. Physically sound estimates of the parameters of an isothermal kernel can be obtained under the assumption that it consists mainly of negative daemons. Proton decay maintains a high temperature of the daemon gas (up to $\sim 10^{11}$–$10^{12}$ K), which makes it physically collisionless and prevents kernel collapse into a black hole.

1. **Introduction**

We start with presenting a remarkable coincidence. Assuming the DM of the Galactic disk to contain daemons, DArk Electric Matter Objects, i.e., electrically charged Planckian elementary black holes (for $Ze = -10$ e, $m = 3.07 \cdot 10^{-5}$ g, $r_g = 1.8 \cdot 10^{-33}$ cm (Drobyshevski, 1998)), the Sun could have captured and accumulated in 4.5 Gyr up to $N \sim 2.4 \cdot 10^{30}$ such particles (Drobyshevski, 1996). Further, as follows from our experiments on detection of negative daemons (Drobyshevski *et al*, 2001), the decay time of a daemon-containing proton is estimated to be $\Delta t_{ex} \sim 10^{-6}$ s. If the time needed for a daemon to capture a proton $\ll \Delta t_{ex}$, and the proton decay liberates $m_p c^2 = 938$ MeV $= 1.5 \cdot 10^{-3}$ erg, the total energy release will be $L_d = 2.4 \cdot 10^{30} \times 1.5 \cdot 10^{-3} / 10^{-6} = 3.6 \cdot 10^{33}$ erg/s, which is practically equal to $L_\odot = 3.86 \cdot 10^{33}$ erg/s. If proton decay provided at least one half of $L_\odot$, this would solve the problem of the deficiency of solar electron neutrinos, even without invoking their oscillations. Therefore, we accept for further estimates $N_- \sim 10^{30}$. This figure correlates with the lower bound on the amount of DM matter in the Galactic disk suggested by Bahcall *et al* (1992), as well as with the present-day general trend to somewhat reduce the amount of DM in the disk (see, e.g., Kuijken, 1999).

We showed earlier that the daemon-assisted catalysis of proton fusion by the β-decay-controlled reaction $10p \to 2\alpha + 4e^+ + 4\nu_e + 2p$ (+53.5 MeV), with reactions occurring in bursts at intervals of $\sim 0.3 \cdot 10^{-8}$ s, exceeds by about a factor five the $L_\odot$ level (Drobyshevski, 1996). Whence it followed that ~4/5 of the daemons would be poisoned by capture of heavy nuclei and, thus, would not take part in the catalysis. This led to the conclusion that the daemon recovers its catalytic properties in $t_{ex} \sim 10^{-7}-10^{-6}$ s as a result of the decay of daemon-containing protons in the captured heavy nucleus (Ne, Fe, etc.) (Drobyshevski, 2000a,b).

For $Z = 10$, the ground state of the daemon is located inside the captured nucleus, and for $Z_n \geq 24/Z$, even inside one of the protons of the nucleus. The daemon binding energy in a nucleus is, in these conditions, $W \approx 1.8ZZ_nA^{-1/3}$ MeV (Cahn and Glashow, 1981). For instance, for a Fe nucleus $W \approx 120$ MeV, which, in the case of its capture, would result in a rapid evaporation of about 15 of its nucleons, including their clusters ($\alpha$–particles etc.), followed by the decay of protons, one after another.

Based on the expected emission of nucleons (and of electrons from the atomic shells), as well as on the fast decay of daemon-containing nucleons, a process that reduces gradually the $Z_n$ of the nuclear remnant to $Z_n < Z$ and makes possible capture of a new nucleus with $Z_n > Z$, we constructed a simple ZnS(Ag) scintillation detector of low-velocity daemons ($V \leqslant 30-50$ km/s) which are members of the Solar system. The instrument detected a flux of highly penetrating particles with the expected properties (Drobyshevski, 2000c). The flux suffers strong seasonal variations. We detected also objects populating contracting geocentric orbits with perigees inside the Earth (Drobyshevski *et al*, 2001).

An analysis of experimental data permitted us to estimate also the time of consecutive decay of each daemon-containing proton in the nuclear remnant as $\Delta t_{ex} \sim 10^{-6}$ s, which exceeds noticeably the figure cited before, $t_{ex} \sim 10^{-7}-10^{-6}$ s, which is the time until the daemon frees itself completely of a nucleus of the type of Ne or Fe captured by it.

This contradiction increases the real value of $t_{ex}$ to $\sim 10^{-5}$ s and more, thus casting doubt on the scenario based on the daemon-assisted catalysis of proton fusion in the Sun. This scenario contained, besides the possible catalyst poisoning by heavy nuclei, a few more weak points in the reasoning (Drobyshevski, 1996), which could hopefully be removed in the future by a more careful analysis. Here they are:

(a) A burst of fusion reactions in the shell surrounding the daemon, whose protons are all in stable levels, releases an energy $\approx 53.5$ MeV (see above). As follows from the momentum conservation law, most of the energy should be carried away, as a rule, by the lightest of the reaction products, i.e., $\nu_e$, $e^+$, and p. Therefore, it is hard to expect that both $\alpha$–particles formed in the process could leave easily the daemon surroundings. A daemon residing inside the $\alpha$–particle in the ground state has a binding energy $W \approx 22$ MeV, and, therefore, one $\alpha$–particle could be expected to remain bound to the daemon. The rapid capture (<<1 μs) by this system of new protons will initiate formation of new unstable compound nuclei of the type of $^5$Li, $^6$B, etc., with an energy release hardly high enough to remove the daemon's poisoning by ejecting a nuclear cluster of the type of an $\alpha$–particle or a heavier nucleus from the vicinity of the daemon. This leads us to an apparently inevitable conclusion that a major part of the energy produced by daemons in the Sun is liberated not through proton fusion catalysis but rather through proton decay in the compound nucleus with $Z_n \leq 10$, which forms or resides in a more or less stable state in the vicinity of the daemon. Note that the mesons, which are apparently emitted in proton decay, interact with the nucleons of this nucleus so weakly that, instead of giving rise to its disintegration (Ashery *et al*, 1981), as a rule carry away their energy and decay in the surrounding medium with production of neutrinos of different flavors.

(b) Another problem is the settling down of daemons to the Sun's center. This settling down was traditionally assumed to create three difficulties, namely: (*i*) the fusion of daemons transforms them to conventional matter (Markov, 1966), the process releasing negligible heat in the case of the Sun (~$10^{-5} L_\odot$); (*ii*) the daemons merge to form one black hole (~$3\cdot10^{25}$ g); the black-hole scenario was analyzed more than once (Hawking, 1971; Clayton *et al*, 1975; Markovic, 1995) and, thus far, has not got observational support, because it does not solve any problem; (*iii*) while solar energetics are supported to some extent by daemon-assisted catalysis of proton fusion, hydrogen (and other light elements) present in a small volume near the center would transform very soon to nuclei with $Z_n > Z$, and the catalytic reactions of fusion would stop.

In an attempt to avoid the settling down of daemons to the Sun's center, a process that appeared unfavorable, we pointed out the possibility of speeding on fast daemons ($V \sim 1000$ km/s) in bursts of fusion reactions, which would occur primarily on their trailing side if the captured proton shell was somehow deformed or oriented by the resistance of the surrounding matter (Drobyshevski, 1996). Straightforward estimates of this possibility show it, however, to be not efficient enough in the solar conditions.

We are going to study below the consequences of the daemons' settling to the Sun's center and see that this process is by far not so unfavorable as this would seem at first glance. The daemon kernel thus formed is capable of providing an appreciable part of the Sun's luminosity not by catalyzing the fusion reactions but rather through daemon-stimulated proton decay. This process, as well as the recently discovered indications of possible neutrino oscillations (to be more rigorous, the "indication of a nonelectron flavor component in the solar neutrino flux") (Ahmad *et al*, 2001), reduces considerably the acuteness of the solar neutrino deficiency problem, because now there is ~one neutrino per ~1 GeV of released energy, rather than ~10 MeV, as was the case with hydrogen fusion.

By way of introduction to the problem, we are going to remind the reader in Sec. 2 of our earlier conclusions (Drobyshevski, 2001, 2002) concerning the properties of the Earth's daemon kernel, which provides the lacking ~10–20 TW of heat, as well as the flow of $^3$He and other light gases emanating from the interior of our planet.

Next (Sec. 3) we shall turn to the Sun. The interaction of the solar kernel with surrounding matter, primarily, the hydrogen plasma, and not iron, as is the case with the Earth, has its specific features. Energy liberation in the kernel extends not through a thin surface layer but over a substantial part of its volume. However, an attempt at understanding the processes occurring in the kernel made up of equal numbers of negative and positive daemons (Sec. 4) yields an unphysical result for the daemon temperature, namely, ~$10^{14}$ K! To be able to construct physically noncontradictory models, one will have to admit a nearly complete absence of positive daemons (Secs. 5 and 6).

In the final Section 7, we note that an analysis of the possible consequences of a daemon kernel formation in the Sun and of the daemon-stimulated proton decays in it, leads one to a conclusion that these reactions play an important part in the total energy balance of the Sun and emission of different flavor neutrinos, as well as that the kernel contains apparently only negative daemons. A desirability is pointed out of an integrated study of daemon kernels in a large variety of celestial objects, starting with the Earth and ending with the pulsars and galaxies.

## 2. The Daemon Kernel of the Earth

A situation similar in many respects to the Solar case is realized inside the Earth and other planets. As follows from measurements of the daemon flux through our detector and of the rate of their slowing down in crossing the Earth, the latter accumulated $N_- \sim 10^{23}$ negative daemons in 4.5 Gyr of its existence (Drobyshevski, 2000d, 2001, 2002).

If we refuse to be frightened by the above difficulties and follow a straightforward logic, we may admit the existence inside the Earth of a kernel of daemons which interact with iron nuclei to liberate a certain energy and to produce isotopes of light gases like $^2$H, $^3$H, $^3$He, $^4$He etc. Indeed, judging from many available studies, about 20 TW of energy (from the total flux $\cong$ 40 TW) emanate from the Earth's interior. The origin of this energy excess, besides the flux of light rare gases, remains unclear to date (Calderwood, 2001). Intense efforts undertaken in the recent decades to unravel its nature still have not met with success (if we disregard notoriously *ad hoc* hypotheses) (Helfrich and Wood, 2001; Porcelly and Halliday, 2001). In accordance with our measurements, we assumed this source (its intensity was accepted for the purpose of estimation as $Q = 10$ TW $= 10^{20}$ erg/s) to be consecutive disintegration of the protons in Fe nuclei captured by the kernel daemons from the surrounding material of the Earth's inner iron core. The decay time of a daemon-containing proton involving liberation of energy $m_p c^2 = 938$ MeV is, as already mentioned, $\Delta t_{ex} \sim 1$ μs (Drobyshevski *et al*, 2001). The daemon is capable of capturing another iron nucleus after the charge of the previous nuclear remnant has dropped as the result of proton disintegration to $Z_n = Z - 1 = 9$ (i.e., the remnant corresponds to the fluorine nucleus with an excess of neutrons, so that in a few seconds it will transform to the Ne nucleus). This recapture is accompanied by a release of energy $\Delta W \approx 18\times(26\times56^{-1/3} - 9\times20^{-1/3}) = 63$ MeV, enough to evaporate 9–10 nucleons (and their clusters of the type $^2$H, $^3$H, $^3$He, $^4$He etc.) from the newly captured nucleus in $\sim 10^{-8}$ s. After disintegration of $\sim$10–11 protons in this nuclear remnant in $t_{ex} \approx 10$–11 μs, the recapture may again occur. If the recapture of nuclei by a daemon and disintegration of their protons take place continuously, such a daemon generates an energy $q \approx 1.5 \cdot 10^3$ erg/s.

To gain an idea of the possible parameters of the kernel and of the daemon plasma in it, we accepted that (Drobyshevski, 2001, 2002):

1) The kernel contains also an equal number of positive daemons ($N_+ = N_- \approx 3 \cdot 10^{23}$), which gives $M_k = 1.8 \cdot 10^{19}$ g for its total mass;
2) The gas kinetic pressure $p_s = n_s kT_k$ of the daemon plasma at the kernel surface is equal to the pressure at the Earth's center $p_c = 3.7 \cdot 10^{12}$ dyne/cm$^2$ (i.e., we neglect the pressure of the other plasma components, such as the pions, positrons, electromagnetic radiation etc.);
3) The density of the surrounding material, i.e., iron, is $r_c = 13$ g/cm$^3$, and its temperature, $T_c = 10^4$ K [in present-day models of the Earth, $T_c = (5.5$–$6.5)\cdot 10^3$ K]; thus, in this stage we practically neglect the action of the kernel on the structure of the adjoining central zone of the Earth's inner core;
4) Iron nuclei ($Z_n = Z_{Fe} = 26$) diffuse from outside into the kernel with a diffusion coefficient $D = l_{Fe}V_{Fe}/3$. Here $V_{Fe} = (3kT_c/m_{Fe})^{1/2}$, and the mean free path $l_{Fe}$ is determined by the Coulomb interaction with near-surface daemons (with due account of the nuclear remnants with $Z_{eff} \approx 5$ captured by them) (e.g. Alfvén and Fälthammar, 1967):

$$l_{Fe} = (3kT_c)^2/(8\pi \ln L e^4 Z_{Fe}^2 Z_{eff}^2 n_s) \tag{1}$$

where $\ln L \approx 10$ is the Coulomb logarithm;

5) The kernel is approximated by a self-gravitating isothermal gas sphere, which material is strongly concentrated toward the center. Within most of its volume, $r \approx r_s R_k^2/R^2$, whence $M(R) \approx M_k R/R_k$ and $r_m = 3M_k/4\pi R_k^3 \approx 3r_s$ (Chandrasekhar, 1939). The concentration of material toward the center is characterized by the ratio of the pressure at the center, which can be only higher than $p_0 = (3GM^2)/(8\pi R_k^4)$ at the center of a uniform sphere (Eddington, 1926), to that at the outer boundary $p_s$ (we shall see that $p_0 >> p_s$). The isothermality is accounted for both by the major energy release occurring near the kernel surface and by the mean free path of a daemon being, as a rule, comparable to $R_k$;

6) The characteristic time scale governing the diffusion of Fe nuclei into the kernel is the time $t_{ex} = 10\text{--}11$ μs between consecutive Fe capture events. For $n_{Fe} >> n_s$, a Fe nucleus diffusing into the kernel penetrates to the depth where it is certain to meet a free *negative* daemon, in a time $t \approx 2t_{ex}n_{Fe}/n_s$. In this case, the depth of diffusion (and of energy release) will be $l = (Dt)^{1/2}$, and for the liberated energy we can write

$$Q = 4\pi R_k^2 \cdot l \cdot q \cdot (n_s/2) \tag{2}$$

Taking $n_s \cong n_m/3 = (N_- + N_+)/4\pi R_k^3$, $p_s = p_c$, and substituting the above relations for $l$, $D$, and $l_{Fe}$, we obtain the following dependence of $R_k$ on $Q$ and other parameters:

$$R_k = (3\ln L e^4 Z_{Fe}^2 Z_{eff}^2 m_{Fe}^{1/2})^{1/4} \cdot [\pi n_{Fe} t_{ex}(3kT_c)^{5/2}]^{-1/4} \cdot (Q/q)^{1/2} \tag{3}$$

Substituting now $T_c = 10^4$ K, $n_{Fe} = r_c/m_{Fe} = 1.4 \cdot 10^{23}$ cm$^{-3}$, $Z_{Fe} = 26$, $Z_{eff} = 5$, and $q = 1.5 \cdot 10^3$ erg/s, we obtain $R_k = 1.6 \cdot 10^{-10} Q^{1/2}$ cm (it does not depend on $M_k$), and, accordingly, for $Q = 10^{20}$ erg/sec, $R_k = 1.6$ cm, $r_s = 0.35 \cdot 10^{18}$ g/cm$^3$, $l = 2 \cdot 10^{-7}$ cm, $T_k = 2.4 \cdot 10^6$ K, which yields for the daemon thermal velocity $V_d \approx 60$ μm/s.

For the kernel parameters obtained, the daemon mean free path for Coulomb collisions is $\sim 10^{-5}$ cm, but the time in which one daemon encounters another in at least one physical collision for the cross section $\pi r_g^2$ is in excess of the age of the Universe. In this sense, the daemon plasma of the kernel is collisionless.

## 3. The Daemon Kernel of the Sun. Main Characteristics and Assumptions

The structure of the kernel of the Sun and the main processes occurring in it should differ from those of the Earth for at least three reasons:

1) The total number of daemons, both negative and, presumably, positive (for $N_- = N_+$) may be as high as $N = N_- + N_+ \approx 2.4 \cdot 10^{30}$ (Drobyshevski, 1996), a figure substantially larger than the one derived from our experiments for the Earth ($N_- \approx 3 \cdot 10^{23}$). In view of the current trend toward decreasing the amount of DM in the Galactic disk (Kuijken, 1999), this figure should possibly be reduced, say, to $N_- \approx 10^{30}$. Whence it follows that the daemon-assisted fraction of the total Sun luminosity can be only $L_d \leqslant N_- \cdot q = 1.5 \cdot 10^{33}$ erg/s.
2) In contrast to the Earth, where the kernel lies inside the iron core, the material at the center of the Sun resides in the state of ideal gas.
3) Because the particles diffusing into the solar kernel from the outside are primarily protons, $Z_n = Z_p = 1$, and $t_{ex} = \Delta t_{ex} \approx 10^{-6}$ s. Therefore, as soon as one of the ten protons captured by a daemon has been digested, a new proton can be captured. Thus, the effective charge $Z_{eff}$ of a negative daemon in the outer active zone of the kernel has, as a rule, two values, 0 or 1.

In view of the above, we shall use a simplified model of the Sun in order to make the calculations less complicated and to gain at least a very general idea of the characteristics of the solar kernel. We shall assume the Sun to consist of hydrogen with the parameters at the center close to those accepted in zero-age Sun models (e.g., Turck-Chiéze *et al*, 1988), namely, $\boldsymbol{r}_{co} = 90$ g/cm$^3$, $T_{co} = 1.5 \cdot 10^7$ K, and $p_{co} = 2.23 \cdot 10^{17}$ dyne/cm$^2$. Within $R \sim 0.001 R_\odot$ from the center, the Sun's density is practically constant. A comparison of the gravitation in a standard Sun with that developed by the kernel shows them to become comparable at a distance $R_x \cong (3M_k/4\pi \boldsymbol{r}_{co})^{1/3} = 0.6 \cdot 10^{-3} R_\odot$. If ~1/3–1/2 of all the energy is liberated at the Sun's center, the whole near-center zone will be in a state of convection, i.e., have a quasi-adiabatic structure described by the equation of state $p = \boldsymbol{k}\boldsymbol{r}^{\boldsymbol{g}}$. For $p = p_{co}$, $\boldsymbol{r} = \boldsymbol{r}_{co}$, and $\boldsymbol{g} = 5/3$, we have $\boldsymbol{k} = 1.23 \cdot 10^{14}$ cm$^4 \cdot$g$^{-2/3} \cdot$sec$^{-2}$.

Integration of the equation of hydrostatic equilibrium which takes into account only the kernel gravitation yields for the hydrogen density at the kernel surface, $R = R_k \ll R_x$,

$$\boldsymbol{r}_{ps}^{\boldsymbol{g}-1} = \boldsymbol{r}_{co}^{\boldsymbol{g}-1} + (GM_k/\boldsymbol{k})\cdot(\boldsymbol{g}-1)/\boldsymbol{g}R_k \qquad (4)$$

or

$$\boldsymbol{r}_{ps}^{2/3} = \boldsymbol{r}_{co}^{2/3} + 2GM_k/5\boldsymbol{k}R_k = a + bR_k^{-1} = \boldsymbol{Q}, \qquad (5)$$

where $a = 20.1$ g$^{2/3}$cm$^{-2}$, $b = 1.33 \cdot 10^4$ g$^{2/3}$cm$^{-1}$ for $M_k = 6.14 \cdot 10^{25}$ g (or $b = 0.8 \cdot 10^4$ g$^{2/3}$cm$^{-1}$ for $M_k = 3.7 \cdot 10^{25}$ g, see below in Sec. 5).

The solar plasma temperature at the kernel's surface

$$kT_{ps} = m_p \boldsymbol{k}(a + bR_k^{-1})/2 = m_p \boldsymbol{k}\boldsymbol{Q}/2 \qquad (6)$$

Consider now two variants of the Sun kernel model similar to the way used with the kernel of the Earth.

## 4. Kernel of Negative and Positive Daemons

We shall start, as before, with the energy balance, and assume for definiteness $Q = L_d \approx 10^{33}$ erg/s and $N = N_+ + N_- = 2 \cdot 10^{30}$. Thus, we assume that, in contrast to the Earth, roughly about one half of all negative daemons are involved in energy liberation, and, therefore, the layer thickness $l$ to which protons have to diffuse into the isothermal kernel should make up a sizable fraction of its radius. For $l \approx R_k/2$, half of the kernel mass is involved; at this depth the concentration $n_d \approx 4n_s$. Therefore, we shall consider the parameters of energy release for an average depth ~$l/2 \sim R_k/4 - R_k/3$, where the negative daemon concentration $n_-(= n_+) = n_d/2 \approx n_s$. Thus, although now we are using the approximation of a thick energy-releasing spherical layer, we retain the quasi-planar approach.

In these conditions, the energy balance can be written

$$Q = 4\pi R_k^2 \cdot l \cdot q \cdot n_s, \qquad (7)$$

where $n_s = N/4\pi R_k^3$, and $l = [(V_{ns} \boldsymbol{l}_p \boldsymbol{t})/3]^{1/2}$ with

$$\boldsymbol{l}_p = (3kT_{ps})^2/(8\pi \ln \boldsymbol{L} e^4 Z_p^2 Z_{eff}^2 n_s) \tag{8}$$

and $\boldsymbol{t} = \boldsymbol{t}_{ex} n_{ps}/n_s$. Here $Z_p = 1$, and $Z_{eff} = 10$ because of the presence of positive daemons with a concentration $n_+ \approx n_s$, whose charge is not compensated by electrons because of ionization at a high $T_{ps}$.

Using the above adiabatic relations (4)–(6) connecting $r_{ps}$, $T_{ps}$, and $p_{ps}$ in the kernel gravitation field as functions of $R_k$, we find from Eqs. (7) and (8)

$$R_k = \{[(2/3)^{3/2}(\ln \boldsymbol{L} e^4 Z_p^2 Z_{eff}^2)/(\pi m_p \boldsymbol{t}_{ex} \boldsymbol{k}^{5/2})]^{1/4} (Q/q)^{1/2} - (2GM_k/5\boldsymbol{k})\} \boldsymbol{r}_{co}^{-2/3} \tag{9}$$

whence $R_k = 2.5 \cdot 10^3$ cm. The pressure of the solar material at this radius $p_{ps} = 4 \cdot 10^{17}$ dyne/cm$^2$ for $r_{ps} = 128$ g/cm$^3$ and $T_{ps} = 1.9 \cdot 10^7$ K. To balance $p_{ps}$, the kernel daemons must have the temperature $T_k \sim 3 \cdot 10^{14}$ K, which is substantially higher than the proton rest mass ($2m_p c^2/3k \sim 7.2 \cdot 10^{12}$ K). This situation has apparently no physical sense. It results actually from our requirement that the proton decay initiated by negative daemons provide a sizable fraction of $L_\odot$. In order for the protons to penetrate into the kernel to a depth large enough to be acted upon by an appreciable part of daemons, the volume concentration of the daemons must be fairly low, i.e., the kernel must be loose and have a large radius. It is this that requires the unreasonably high daemon temperature.

5. **Kernel of Negative Daemons. Inclusion of their Thermal Ionization**

The situation changes radically if the daemons in the Sun are only (or predominantly) of the negative species. There is nothing particular in this assumption, especially if we recall that there are practically no antiparticles in our Universe. The excess electrical charge of daemons in the kernel can be compensated by both protons and positrons, which are produced in the decay of the latter.

We take again $Q = 10^{33}$ erg/sec, but now $N \equiv N_- = 1.2 \cdot 10^{30}$, and $M_k = 3.7 \cdot 10^{25}$ g. The parameters of the solar material near the kernel vary adiabatically in accordance with Eqs. (4)–(6), but with $b = 0.8 \cdot 10^4$ g$^{2/3}$·cm$^{-1}$. We retain the thick diffusion layer approximation, in which the protons have to diffuse fairly deep into the kernel, where $n_d \approx 2n_s$ (i.e., $l \sim R_k/2$). We adopt that during the time interval $\boldsymbol{t} \approx \Delta \boldsymbol{t}_{ex} n_p/2n_s$ protons ($Z_p = 1$) diffuse through a mixture of daemons whose charge is fully compensated by ten captured protons ($Z_{eff} = 0$; it is an analog of the antineon atom with the first ionization potential $\chi = 21.56 \times 1837$ eV = 39.6 keV) and thermally single-ionized antineon atoms ($Z_{eff} = Z_\alpha = 1$) (positive daemons, if present, are fully ionized, so that consideration in their presence of thermal ionization of negative daemons, which we are going to do below, would hardly affect $\boldsymbol{l}_p$ in the estimates we made in the preceding Sec. 4).

Then the proton mean free path in a daemon plasma will be

$$\boldsymbol{l}_p = (3kT_{ps})^2/(8\pi \ln \boldsymbol{L} e^4 Z_p^2 Z_\alpha^2 2n_s \boldsymbol{a}), \tag{10}$$

where the degree of ionization $a$ is determined from the Saha equation (e.g., Allen, 1973) for $T = T_{ps}$, with due inclusion of the fact that the pressure of the proton component is $p_{ps}/2$:

$$a = 2p_{ps}^{-1}[(U_1/U_0)^2 2(2\pi m_p)^{3/2}(kT_{ps})^{5/2}h^{-3}\times\exp(-\chi/kT_{ps})] \quad (11)$$

or, because $T_{ps} = (\mathbf{k}m_p/2k)\mathbf{Q}$ and $p_{ps} = \mathbf{k}\mathbf{Q}^{5/2}$

$$a = 2\pi^{3/2}(U_1/U_0)m_p^4 h^{-3}\mathbf{k}^{3/2}\times\exp(-2\chi/\mathbf{k}m_p\mathbf{Q}) \quad (12)$$

For neon, the ratio of the distribution functions $U_1/U_0 = 5.5$ (Allen, 1973); here h is the Planck constant.

Now we can write the energy balance

$$Q = 4\pi R_k^2 q \mathcal{X} 2 n_s \quad (13)$$

in the form

$$Q = [(3/4)(3/2\pi)^{1/2}(h/m_p)^3(U_0/U_1)(\mathbf{t}_{ex}\mathbf{k}/\ln \mathbf{L})]^{1/2}(qR_k^2\mathbf{Q}^2/e^2 Z_p Z_\alpha)\times\exp(\chi/m_p\mathbf{k}\mathbf{Q}) \quad (14)$$

Its solution yields $R_k = 400$ cm, whence $n_s = 1.5 \cdot 10^{21}$ cm$^{-3}$, $T_{ps} = 3 \cdot 10^7$ K, $p_{ps} = 1.25 \cdot 10^{18}$ dyne/cm$^2$, $a = 0.015$, and $T_k = 6.1 \cdot 10^{12}$ K.

Now the daemon temperature in the kernel appears more reasonable (in any case, $kT_k < 2m_p c^2/3$; we may recall that the difference between $T_k$ and $T_{ps}$ seems only natural because of the huge difference between $m$ and $m_p$, with the energy being released predominantly in one of the plasma components – a phenomenon well known, e.g., from the glow gas discharge). At this $T_k$, daemons move with a velocity of only $V_d = 9$ cm/s. The value of $T_k$ can be reduced by a few times (possibly, by an order of magnitude; see also the next Sec. 6) if we take into account that the pressure of the solar matter is compensated not in a thin surface layer of the kernel but rather by a pressure gradient across a thick layer $l \sim R_k$, and that electrons of the outer plasma cross possibly nearly all of the kernel. We have also disregarded the heating of the proton component inside the kernel, a factor that increases its diffusibility, and, hence, reduces $R_k$ and $T_k$. At the same time, however, the presence of He and heavier nuclei in the solar material may act in the opposite sense.

6. **Kernel of Negative Daemons. The Case of their Ionization Due to Proton Decay**

In the preceding Sec. 5 we have considered thermal ionization of antineon atoms, i.e., formations in which a negative daemon ($Z = 10$) is surrounded by ten nonfused protons moving around it. The lifetime of this system until proton fusion reactions flare up in it $\mathbf{t}_{pp} \sim 3 \cdot 10^{-8}$ s (Drobyshevski, 1996). We have, however, pointed out in Introduction that the situation, in which the daemon is embedded in a compound nucleus representing a product of numerous previous proton fusion events in its vicinity, is a more probable case. We accept $Z_n = Z$ for the nuclear charge. Obviously enough, for a binding energy $W \approx 68$ MeV thermal ionization, i.e., detachment of this nucleus from the daemon at $T_{ps}$ is impossible. Ionization occurs once in $\Delta \mathbf{t}_{ex} \approx 10^{-6}$ s only due to daemon-stimulated proton decay in the nucleus itself. The lifetime of the system in the ionized state ($Z_{eff} = 1$), $\mathbf{t}_{cap}$, is determined by the capture of a new proton from the surrounding matter. The capture may occur when two protons collide within a sphere of radius $r_{cap}$ near the $Z_{eff}$ daemon. The radius $r_{cap}$ is determined by the condition

$$Z_p Z_{eff} e^2 / r_{cap} \approx 3kT_{ps}/2, \text{ i.e. } r_{cap} = 2Z_p Z_{eff} e^2 / 3kT_{ps} \tag{15}$$

The proton Coulomb collision frequency per unit volume for $Z_{eff} = Z_p = 1$ can be written as (Alfvén and Fälthammar, 1967)

$$\boldsymbol{n}_{pp} = (0.714 \cdot 8\pi \ln \boldsymbol{L} e^4 n_{ps}^2)/[m_p^{1/2}(3kT_{ps})^{3/2}] \tag{16}$$

whence for the time between collisions of two protons in a sphere of radius $r_{cap}$, i.e., the capture time, we obtain

$$\boldsymbol{t}_{cap} = \boldsymbol{x} \cdot [3m_p^{1/2}(3kT_{ps})^{9/2}]/(0.714 \cdot 256\pi^2 \cdot \ln \boldsymbol{L} \cdot e^{10} \cdot n_{ps}^2) \tag{17}$$

Here the factor $\boldsymbol{x}$ accounts for the fact that not each proton collision by far in a sphere with $r = r_{cap}$ will result in a proton capture by a daemon with $Z_{eff} = 1$. Assuming now $Z_a = 1$ and $\boldsymbol{a} = \boldsymbol{t}_{cap}/\Delta \boldsymbol{t}_{ex}$ in Eq. (10), Eq. (13) for the energy balance reduces to

$$Q = (32\pi/9) \cdot (2 \cdot 0.714\pi/\boldsymbol{x})^{1/2} (e/m_p)^3 (\boldsymbol{t}_{ex}/\boldsymbol{k}) \cdot q \cdot \boldsymbol{Q}^{5/4} R_k^2 \tag{18}$$

Whence for $\boldsymbol{x} = 10$ we obtain only $R_k = 11$ cm.

The reason for such a small value of $R_k$ is the negligibly small effective cross section of interaction of protons diffusing into the kernel with daemons, with the charge of most of them being compensated by the charge of the compound nucleus in which it resides.

For $R_k = 11$ cm, $\boldsymbol{Q} = 740$ g$^{2/3}$·cm$^{-2}$, whence $\boldsymbol{r}_{ps} = 2 \cdot 10^4$ g/cm$^3$, $n_{ps} = 1.2 \cdot 10^{28}$ cm$^{-3}$, $p_{ps} = 1.83 \cdot 10^{21}$ dyne/cm$^2$, $T_{ps} = 0.55 \cdot 10^9$ K, $n_s = 0.7 \cdot 10^{26}$ cm$^{-3}$, $T_k = 1.9 \cdot 10^{11}$ K, and $V_d \approx 1.6$ cm/s. Hopefully, a careful inclusion of the energy exchange between the daemon component and solar plasma will show that (*i*) the kernel is fairly transparent for the diffusion of radiation and protons into it (so that the daemon gas will not be required to compensate all of the external pressure), (*ii*) the energy released in the kernel $L_d \approx Nq$, and (*iii*) its radius $R_k$ will somewhat exceed the value found here.

## 7. Conclusion

The above reasoning suggests the following two important conclusions:

1) The existence of daemons and their accumulation at the Sun's center is capable of accounting for a sizable fraction of solar luminosity through daemon-stimulated proton decay, with the corresponding decrease of the electron neutrino yield. Now the exact contribution of the daemon luminosity remains unclear, if for no other reason than the inevitably rough estimate of the low-velocity daemons captured by the Sun. Their spatial concentration in the Galactic disk and, hence, the flux onto the Sun, may undergo noticeable fluctuations, say, as a result of modulation by the spiral density waves. It is hoped that a definite opinion concerning number of daemons in the Sun, their contribution into $L_\odot$, and possible modes of the daemon-containing proton decay will be composed basing on results of experiments detecting fluxes of solar neutrinos of different flavors (e.g. Ahmad *et al*, 2001). The kernel existence, to say the least, does not contradict to different helioseismological data which confirm the Standard Solar Model (SSM) and provide fairly self-consistent results down to depth $R \gtrsim 0.1\text{-}0.15R_\odot$ (Bahcall *et al*, 1998), i.e. down to the region inside which ~$2/3L_\odot$ is liberated. Moreover, some helioseismological data reveal an increase of the sound speed (and/or smaller density if compared with the SSM) in the Solar core at $R < 0.1\text{-}0.15R_\odot$ (e.g., Gough *et al*, 1996) that can be interpreted as favoring the ideas developed.

2) It appears that only negative daemons exist in our Universe. Additional light may be shed on this problem by comprehensive calculations of evolutionary models of the Sun, stars, planets, and other bodies, made with due account of the gradual growth of the kernel and in comparison with neutrino and more advanced daemon experiments. Development of such models requires taking into account radically new processes at diverse physical levels and is, therefore, a much more complex problem than, say, calculations of stellar models with inclusion of only thermonuclear processes in quasi-equilibrium thermodynamic conditions (we may recall here the large difference between $T_k$ and $T_p$).

The absence of positive daemons should not affect as critically the parameters of the Earth's kernel found earlier, because here $Z_{eff}$ remains large ($Z_{eff} \approx 5$) as a result of the negative daemons capturing multi-charged iron nuclei. Clearly, studying the kernels of the Earth, planets, pulsars etc. should be pursued in close correlation with the investigation of the kernels in the Sun and the stars.

It might seem at first glance that the daemon kernel is practically an analog of the black hole. This is not so, though, because while transforming the surrounding matter to energy it does not increase in mass. Therefore, the process does not progress, and there are no grounds to fear that, say, the Sun could disappear under some conditions to become a black hole.

Despite the huge density of matter in the kernel ($\gtrsim 10^{20}$ g/cm$^3$), the daemons in it constitute actually a collisionless plasma, where the time of at least one encounter of two daemons to a distance ~$3r_g$ in the whole kernel may be measured in months and years. Therefore, in the conditions favorable for energy release in proton disintegration occurring in stars and planets daemon fusion and kernel transformation into a black hole are hardly conceivable. However, in the case of originally much more massive condensations of daemons their relativistic collapse at a certain evolutionary stage should not possibly be excluded (quasars, AGN, etc.?).